\documentclass[preprint,aps,floatfix,showkeys]{revtex4}

\usepackage{graphicx}% Include figure files
\usepackage{dcolumn}% Align table columns on decimal point
\usepackage{bm}% bold math

\nonstopmode
\usepackage{epsfig}
\begin{document}
\title{Revealing molecular orbital gating by transition-voltage spectroscopy\footnote{Dedicated to Professor Horst K\"oppel on the occasion of his 60$^{th}$
birthday}}

\author{Ioan B\^aldea}
\email{ioan.baldea@pci.uni-heidelberg.de}
\altaffiliation[also at ]{National Institute for Lasers,
Plasmas, and Radiation Physics, ISS,
RO-077125 Bucharest, Romania.}
\affiliation{Theoretische Chemie,
Physikalisch-Chemisches Institut, Universit\"{a}t Heidelberg, Im
Neuenheimer Feld 229, D-69120 Heidelberg, Germany}
\date{\today}

$ $ \\[2ex]
\hspace{0cm}

\begin{abstract}
Recently, Song et al [Nature 462, 1039 (2009)] 
employed transition-voltage spectroscopy to demonstrate that 
the energy 
$\varepsilon_H$ of the highest occupied molecular orbital (HOMO) 
of single-molecule transistors can be 
controlled by a gate potential $V_G$. 
To demonstrate the linear dependence $\varepsilon_H-V_G$, the experimental 
data have been interpreted by modeling 
the molecule as an energy barrier 
spanning the spatial source-drain region of 
molecular junctions. Since, as shown in this work, that
crude model cannot quantitatively describe the measured  
$I$-$V$-characteristics, it is important to get further support for 
the linear dependence of $\varepsilon_H$ on $V_G$. 
The results presented here, which have been obtained within a model 
of a point-like molecule, confirm this linear dependence. 
Because the two models rely upon complementary descriptions,
the present results indicate that the interpretation of the 
experimental results as evidence for a gate controlled HOMO 
is sufficiently general.
\end{abstract}
\maketitle

\renewcommand{\textfraction}{0}

%--------------------------------
%%%%%%%%%%%%%%%%%%%%%%%%%%%%%%%%%%%%%%%%%%%%%%%%%%%%%%%%%%%%%%%%%%%%%%%%%
%%%%%%%%%%%%%%%%%%%%%%%%%%%%%%%%%%%%%%%%%%%%%%%%%%%%%%%%%%%%%%%%%%%%%%%%%
\section{Introduction}
\label{sec-introduction}
The practical realization of ever smaller transistors is of paramount 
importance for the future of nano- and molecular-electronic devices. 
Transistors of nanoscopic sizes fabricated by linking a lithographically 
defined semiconducting quantum dot to electrodes \cite{Goldhaber-GordonNature:98} are based on 
the Kondo effect. In even smaller transistors, the current flows through a single 
molecule. Besides the Kondo mechanism, the Coulomb blockade can also be used 
for achieving transistor functions with correlated single molecules 
\cite{park02,Liang:02,Kubatkin:03,Natelson:04}.
For uncorrelated molecules, transistor-like $I$-$V$-characteristics 
can be obtained 
by modifying electrostatically the molecular orbital energies, a mechanism
predicted theoretically \cite{Datta:04} and realized in electrochemical 
environment \cite{Tao:05,Wandlowski:08} as well in a beautiful recent 
experiment \cite{Reed:09}. 
In all these (field-effect) transistors, 
the current $I$ which flows due to a bias $V$ applied between two electrodes  
(source and drain) is controlled by means of the voltage $V_{G}$ applied on 
a third electrode (gate). 

The very detailed characterization of the molecular 
devices of Ref.~\cite{Reed:09} has allowed to convincingly demonstrate
that transistor functions can be achieved by means of single-molecule devices. 
In Sec.~\ref{sec-exp}, the experimental evidence on the molecular orbital gating 
obtained by transition-voltage spectroscopy will be reviewed. Importantly, 
as it will be shown there,
the model used in Ref.~\cite{Reed:09} to describe the active molecules 
cannot quantitatively reproduce the $I$-$V$-curves measured 
experimentally. So, it is important to  
show that the linear dependence of the highest occupied molecular 
orbital (HOMO) energy $\varepsilon_H$
on the gate potential $V_G$, a key feature demonstrating the HOMO-mediated 
transistor function,
does not critically depend on the particular model assumed in Ref.~\cite{Reed:09}. 
To this aim, we shall consider in Sec.~\ref{sec-pcm} a complementary model, 
namely a point-like molecule.
As discussed in Sec.~\ref{sec-discussion}, this model reconfirms the aforementioned 
linear dependence. The conclusion of the present investigation, presented in  
Sec.~\ref{sec-conclusion}, is that the linear dependence 
of $\varepsilon_H$ on $V_G$ holds beyond the barrier model utilized in Ref.~\cite{Reed:09} 
and is therefore more general.
%%%%%%%%%%%%%%%%%%%%%%%%%%%%%%%%%%%%%%%%%%%%%%%%%%%%%%%%%%%%%%%%%%%%%%%%%
\section{Review of the experimental evidence on the molecular orbital gating}
\label{sec-exp}
\begin{figure}[h!]
% \centerline{\hspace*{-0ex}\includegraphics[width=0.3\textwidth,angle=0]{fig_nature_fit_odt.eps}}
\centerline{\hspace*{-0ex}\includegraphics[width=0.8\textwidth,angle=0]{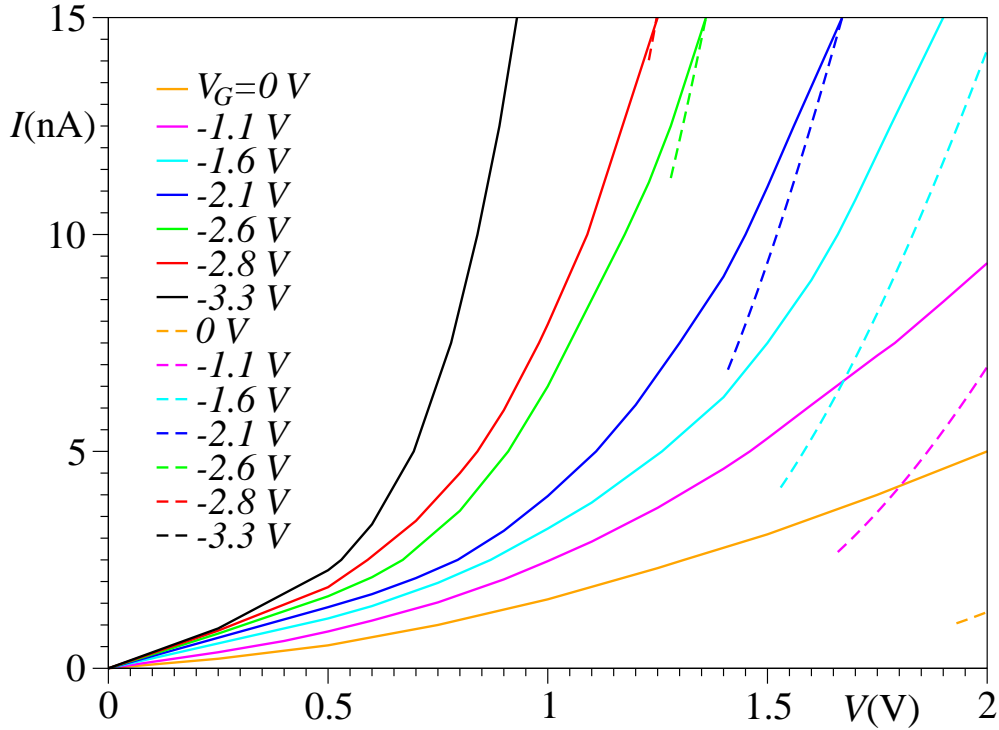}}
\caption{Gate controlled $I$-$V$ characteristics of the Au-ODT-Au molecular 
junctions of Ref.~\cite{Reed:09} for 
several gate potentials $V_{G}$ given in the legend. 
The solid lines have been obtained by extracting the experimental 
data from Fig.~1a of Ref.~\cite{Reed:09} for ODT-based molecular transistors.
The dashed lines, shown for voltages above the transition voltage ($V > V_{t}^{e}$),  
should represent the high voltage approximation [Eq.~(\ref{eq-fn})]
of the experimental curves obtained by modeling the active molecule as a 
continuously 
expanded energy barrier, but, as visible in this figure, this approximation
is very poor. See the main text for details.}
\label{fig:iv-nature}
\end{figure}

The $I$-$V$-curves measured in Ref.~\cite{Reed:09} for single-molecule transistors based 
on 1,8-octanedithiol (ODT) and 1,4-benzenedithiol (BDT) exhibit a sensitive
dependence on the gate potential (cf.~Fig.~\ref{fig:iv-nature}) revealing 
thereby a clear transistor function. However these curves alone do not 
yet provide much insight into the underlying physical mechanism. 
Demonstrating that the current really flows through the molecule under consideration 
and ruling out spurious contributions from impurities and defects 
represent important challenges for molecular electronics altogether.

%%%%%%%%%%%%%%%%%%%%%%%%%%%%%%%%%%%%%%%%%%%%%%%%%%%%%%%%%%%%%%%%%%%%%%
%%%%%%%%%%%%%%%%%%%%%%%%%%%%%%%%%%%%%%%%%%%%%%%%%%%%%%%%%%%%%%%%%%%%%%

A key issue is to provide direct evidence on the electrostatical modulation of the 
energy $\varepsilon_{0} (= \varepsilon_H$ or $\varepsilon_L$) 
of the frontier molecular orbitals,
i.~e.~the energy $\varepsilon_H$ of 
the highest occupied molecular orbital (HOMO; p-type conduction) or 
the energy $\varepsilon_L$ of 
the lowest unoccupied molecular orbital (LUMO; n-type conduction), 
depending on which of these molecular orbitals 
is the closest to the electrode Fermi energy $\varepsilon_F$. (Unless otherwise specified, 
$\varepsilon_F$ is chosen as zero of energy.) 
Is this the case, and does the device act indeed as a single-molecule transistor, 
the gate-controlled molecular energy $\varepsilon_{0}$ should linearly depend on 
the gate potential 
%%%%%%%%%%%%%%%%%%%%%%%%%%%%%%%%%%%%%%%%%%%%%%%%%%%%%%%%%%%%%%%%%%%%%%%%%
\begin{equation}
\label{eq-alpha}
\varepsilon_{0} = \varepsilon_{0}^{0} + \alpha e V_{G} . 
\end{equation}
%%%%%%%%%%%%%%%%%%%%%%%%%%%%%%%%%%%%%%%%%%%%%%%%%%%%%%%%%%%%%%%%%%%%%%%%%
Here $e > 0$ is the elementary charge and $\varepsilon_0^{0}$ the energy of 
the frontier molecular orbital without applied gate voltage.
Similar to other cases 
encountered in nanotransport discussed recently in several papers coauthored 
by K\"oppel 
\cite{Baldea:2009a,Baldea:2010d,Baldea:2010b,Baldea:2010c}, unlike
the gate voltage $V_{G}$, 
neither the level energy $\varepsilon_{0}$ nor the 
conversion factor $\alpha$ are directly accessible experimentally. So, a 
special procedure is required to verify the validity of Eq.~(\ref{eq-alpha}).

To extract the energy  $\varepsilon_{0}$ from the measured $I$-$V$-curves, 
which further allowed to demonstrate the linear dependence  
$\varepsilon_{0} = \varepsilon_{0}(V_{G})$,
Song et al \cite{Reed:09} used the so-called transition-voltage spectroscopy 
\cite{Beebe:06,Choi:08,Chiu:08,Choi:10,Kushmerick:09}. 
This technique exploits the existence of two different 
regimes in the charge transport by tunneling 
through molecular junctions. The distinction between these two mechanisms
was pointed out long time ago by means of semiclassical WKB calculations of the electric 
current between two metallic electrodes separated by a thin insulating film 
\cite{Simmons:63}. The latter was modeled as an energy barrier continuously 
filling the space between electrodes.
%%%%%%%%%%%%%%%%%%%%%%%%%%%%%%%%%%%%%%%%%%%%%%%%%%%%%%%%%%%%%%%%%%%%%%
\begin{figure}[h]
% \centerline{\hspace*{-0ex}\includegraphics[width=0.3\textwidth,angle=0]{RectTrapezTriang.eps}}
\centerline{\hspace*{-0ex}\includegraphics[width=0.8\textwidth,angle=0]{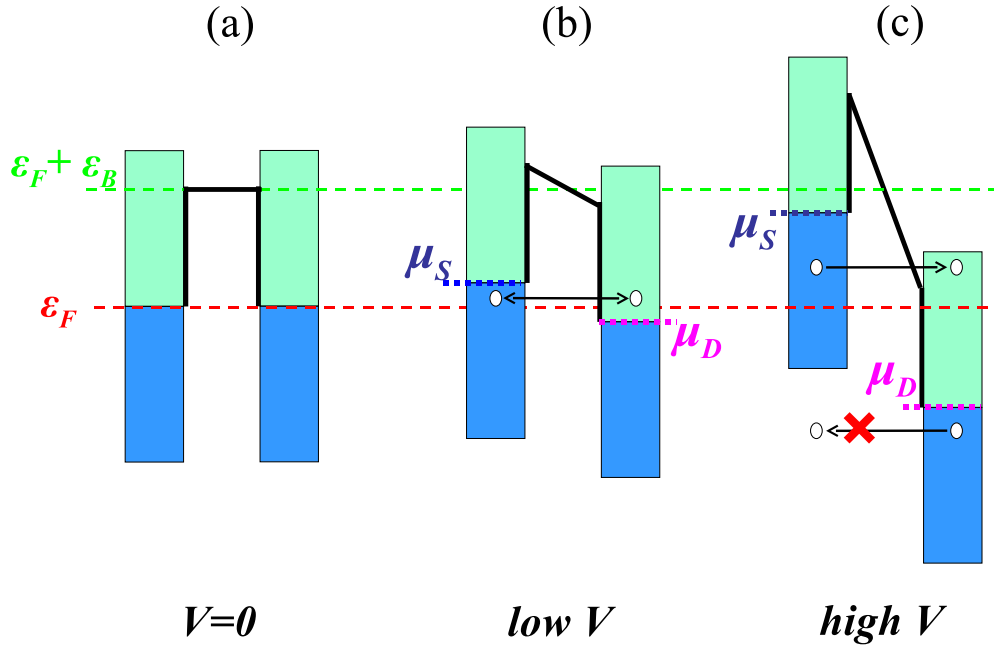}}
\caption{Modeling a molecule spanning the region between the source (S) 
and the drain (D) of a molecular junction by a homogeneous energy barrier continuous 
in space whose shape is (a) rectangular, (b) trapezoidal, or (c) triangular,
depending on the magnitude of 
the source-drain voltage $V$ ($\mu_{S,D} = \varepsilon_F \pm eV/2$). 
The situation depicted here 
(energy barrier for $V=0$ above the electrodes' Fermi energy 
$\varepsilon_F$) corresponds to a LUMO-mediated conduction.}
\label{fig:rect-trapez-triang}
\end{figure}
%%%%%%%%%%%%%%%%%%%%%%%%%%%%%%%%%%%%%%%%%%%%%%%%%%%%%%%%%%%%%%%%%%%%%%

Without a source-drain bias ($V=0$), the barrier can be assumed rectangular,
with a certain height $\varepsilon_B$ (Fig.~\ref{fig:rect-trapez-triang}a). 
A low bias ($e V\ll \varepsilon_B$) changes this rectangular barrier to a 
trapezoidal one 
(Fig.~\ref{fig:rect-trapez-triang}b) 
and yields an ohmic regime 
%%%%%%%%%%%%%%%%%%%%%%%%%%%%%%%%%%%%%%%%%%%%%%%%%%%%%%%%%%%%%%%%%
\begin{equation}
\label{eq-ohm}
I \simeq A V \exp\left(-\gamma \varepsilon_{B}^{1/2}\right) .
\end{equation}
%%%%%%%%%%%%%%%%%%%%%%%%%%%%%%%%%%%%%%%%%%%%%%%%%%%%%%%%%%%%%%%%%
By sufficiently raising the bias ($e V>\varepsilon_B$), 
the barrier becomes triangular. 
If the Fermi level of the drain lies below the bottom of the 
conduction band of the source 
(Fig.~\ref{fig:rect-trapez-triang}c) one arrives at a situation 
that resembles the field emission of electrons from a metallic 
electrode. In this regime (Fowler-Nordheim tunneling), the current 
can be expressed as 
%%%%%%%%%%%%%%%%%%%%%%%%%%%%%%%%%%%%%%%%%%%%%%%%%%%%%%%%%%%%%%%%%
\begin{equation} \label{eq-fn}
I \simeq \left(B V^2/\varepsilon_B\right) 
\exp\left(- \delta \varepsilon_{B}^{3/2}/V \right) .
\end{equation}
%%%%%%%%%%%%%%%%%%%%%%%%%%%%%%%%%%%%%%%%%%%%%%%%%%%%%%%%%%%%%%%%%
Above, $A$, $\gamma$, $B$, and $\delta$ are parameters that do not depend on
$V$ and $\varepsilon_B$.

The transition between these two regimes can be visualized with the aid 
of the so-called Fowler-Nordheim plot, which is the graph representing the 
quantity $\log (I/V^2)$ as a function of $1/V$. 
As shown by Eqs.~(\ref{eq-ohm}) and (\ref{eq-fn}),
$\log (I/V^2)$ increases (logarithmically) with $1/V$
in the ohmic regime
%%%%%%%%%%%%%%%%%%%%%%%%%%%%%%%%%%%%%%%%%%%%%%%%%%%%%%%%%%%%%%%%%
\begin{equation}
\label{eq-log-ohm}
\log (I/V^2)\simeq const + \log (1/V) ,
\end{equation}
%%%%%%%%%%%%%%%%%%%%%%%%%%%%%%%%%%%%%%%%%%%%%%%%%%%%%%%%%%%%%%%%%
while it decreases (linearly) in the Fowler-Nordheim regime
%%%%%%%%%%%%%%%%%%%%%%%%%%%%%%%%%%%%%%%%%%%%%%%%%%%%%%%%%%%%%%%%%
\begin{equation}
\label{eq-log-fn}
\log (I/V^2)\simeq const -\delta \varepsilon_{B}^{3/2} (1/V) .
\end{equation}
%%%%%%%%%%%%%%%%%%%%%%%%%%%%%%%%%%%%%%%%%%%%%%%%%%%%%%%%%%%%%%%%%
In Ref.~\cite{Beebe:06}, 
the crossover between these two regimes has not been 
deduced from a specific analytic dependence $I=I(V)$
but rather claimed in view a simple intuitive argument underlying the 
above analysis. Namely, it was specified by the point where the energy barrier changes from 
the trapezoidal shape to the triangular one. This condition 
defines a
transition voltage $V=V_{t}^{e}$ which can be expressed as \cite{Beebe:06} 
%%%%%%%%%%%%%%%%%%%%%%%%%%%%%%%%%%%%%%%%%%%%%%%%%%%%%%%%%%%%%%%%%
\begin{equation}
\label{eq-Vt}
e V_{t}^{e} = \varepsilon_B .
\end{equation}
%%%%%%%%%%%%%%%%%%%%%%%%%%%%%%%%%%%%%%%%%%%%%%%%%%%%%%%%%%%%%%%%%
Importantly, the transition voltage of Eq.~(\ref{eq-Vt}) depends 
on the barrier height. Therefore
$\varepsilon_B$ can be estimated from $V_{t}^{e}$, because the latter 
can be deduced experimentally from the Fowler-Nordheim plot. 
On this basis, a number of recent experimental studies
\cite{Beebe:06,Choi:08,Chiu:08,Choi:10} 
succeeded to reveal a Fowler-Nordheim transition 
in molecular transport.

Basically, to produce a current through the molecule, 
the electrons have to overcome an 
energy barrier equal to the energy offset between 
the frontier molecular orbital 
($\varepsilon_0 = \varepsilon_H$ or $\varepsilon_L$) 
and the electrode Fermi level $\varepsilon_F$,
i.~e.\ $\varepsilon_B = \varepsilon_{F} - \varepsilon_H$ 
or $\varepsilon_B = \varepsilon_L - \varepsilon_F$.
for p-type or n-type conduction, respectively.
To reveal that the current through the molecules investigated in Ref.~\cite{Reed:09} 
was gated by the HOMO energy, 
transition-voltage spectroscopy turned out to be an essential tool. The more negative
the applied gate potential $V_{G}$, the larger was the measured current through 
the molecular transistors of Ref.~\cite{Reed:09}.
This indicates that the conduction is mediated 
by the HOMO and not by the LUMO; a negative gate potential pushes the HOMO energy 
$\varepsilon_{H}$ upwards, and it becomes closer to the electrode's 
Fermi energy; thus, the energy barrier 
$\varepsilon_B = \varepsilon_F - \varepsilon_{H}$ is reduced. 
Most important, the transition
voltage $V_{t}$ 
deduced experimentally from the minimum of the Fowler-Nordheim plot 
\cite{inflexion,frisbie}
was found to linearly vary with the gate potential $V_{G}$, in agreement with 
Eqs.~(\ref{eq-Vt}) and (\ref{eq-alpha}). The large conversion factors extracted in this way 
($\alpha = +0.25$ and $\alpha=+0.22$ for the ODT and BDT molecules, respectively 
\cite{Reed:09}) 
indicate a particularly strong coupling between the embedded molecules and the gate.

The experimental confirmation of the linear relationship $V_{t} = f(V_{G})$
represents a remarkable evidence of the fact that the current through the
single-molecule transistors proceeds through the gate-controlled HOMO of the 
two molecules (ODT and BDT) investigated in Ref.~\cite{Reed:09}. 
However, further work has to be done for the 
quantitative description of the experimental data of Ref.~\cite{Reed:09}.

One can easily show that Eqs.~(\ref{eq-ohm}) and (\ref{eq-fn})
cannot quantitatively reproduce the experimental $I$-$V$-characteristics.
If Eq.~(\ref{eq-fn}) were applicable, for biases $V > V_{t}^{e}$ 
the current at various gate voltages $V_G$ ($\varepsilon_B$) 
could be described by fitting the parameters $B$ and $\delta$.
To illustrate that this is \emph{not} the case, in Fig.~\ref{fig:iv-nature} 
the values of $B$ and $\delta$ have been adjusted to reproduce  
the points ($I, V$) of coordinates 
$\approx$($15.0$\,nA, $1.67$\,V) and $\approx$($15.0$\,nA, $1.36$\,V)
at the ends of the experimental curves for 
$V_G=-2.1$\,V and $V_G=-2.6$\,V. The curves obtained 
from this fit, which are 
depicted by the dashed lines of Fig.~\ref{fig:iv-nature}, represent 
unsatisfactory approximations of the experimental ones.
%%%%%%%%%%%%%%%%%%%%%%%%%%%%%%%%%%%%%%%%%%%%%%%%%%%%%%%%%%%%%%%%%%%%%%%%%
\section{Fowler-Nordheim plots for a point-like molecule}
\label{sec-pcm}
The field of molecular electric transport has long been plagued by notorious 
quantitative discrepancies between the current measured experimentally and 
computed theoretically. Until reaching the desirable agreement between them,
it is important to demonstrate that the interpretation based on a gate 
controlled HOMO-mediated current holds beyond the model discussed in 
Sec.~\ref{sec-exp}. 

Briefly, the model yielding the key Eq.~(\ref{eq-Vt})
assumes an expanded molecule described by 
an energy barrier with a \emph{continuous} spatial extension, which smoothly 
fills the source-drain space and has 
a position-dependent height linearly interpolated between the source and drain 
chemical potentials $\mu_{S,D}$ (cf.~Fig.~\ref{fig:rect-trapez-triang}). 
This is one extreme model of a molecule embedded into a junction.
At the other extreme, one can model the active molecule as a 
\emph{point}-like level of energy $\varepsilon_0$, from/to which the 
electrons are transferred to/from the electrodes. Let us assume that this point-like molecule 
is located symmetrically between the source and the drain 
\cite{symmetrical-location},
and that the bias is applied 
symmetrically, i.~e., $\mu_{S,D} = \varepsilon_F \pm e V/2$. 
%%%%%%%%%%%%%%%%%%%%%%%%%%%%%%%%%%%%%%%%%%%%%%%%%%%%%%%%%%%%%%%%%%%%%%
\begin{figure}[h]
% \centerline{\hspace*{-0ex}\includegraphics[width=0.3\textwidth,angle=0]{PotentialDrop.eps}}
\centerline{\hspace*{-0ex}\includegraphics[width=0.8\textwidth,angle=0]{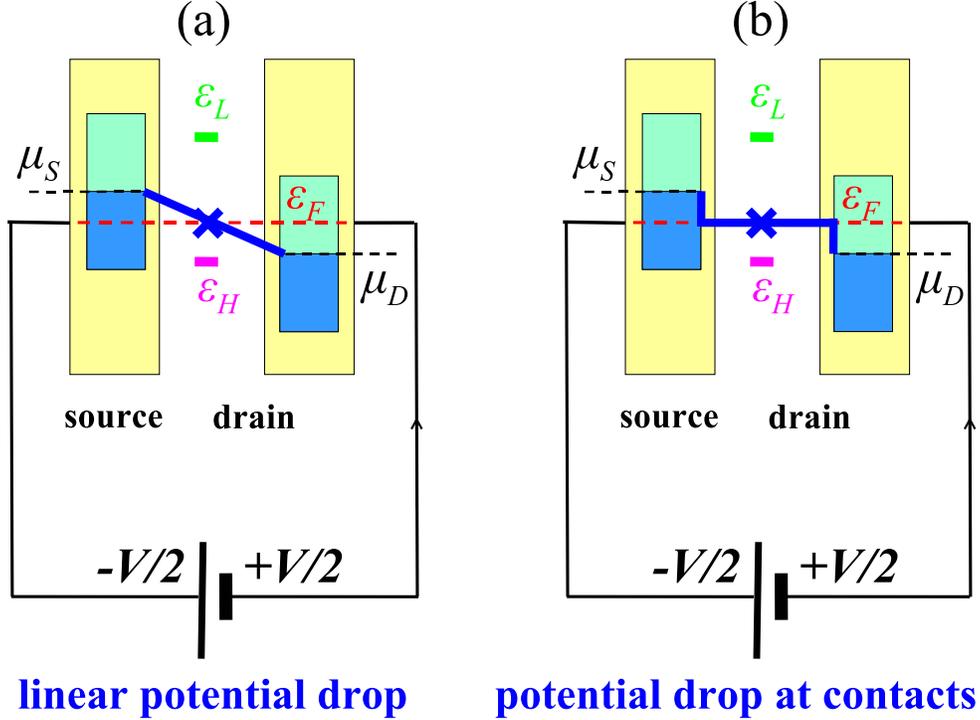}}
\caption{Potential profile (blue line online) 
across the molecular junction: (a) linear variation between 
the source and the drain; (b) sharp potential drop at the two electrodes. 
In both cases, as visualized by the 
symbol $\times$ (blue online), 
the magnitude of the potential drop $V=(\mu_S - \mu_D)/e$ 
does not affect the energy of the HOMO ($\varepsilon_H$) 
and the LUMO ($\varepsilon_L$) of a point-like molecule located 
symmetrically between the source and the drain.}
\label{fig:potential-drop}
\end{figure}
%%%%%%%%%%%%%%%%%%%%%%%%%%%%%%%%%%%%%%%%%%%%%%%%%%%%%%%%%%%%%%%%%%%%%%
Then, whether the potential $V$ varies
linearly between the source and the drain, or it sharply drops near electrodes and remains constant 
practically in the whole region between electrodes, the level energy 
$\varepsilon_0$ does not depend on $V$ (see Fig.~\ref{fig:potential-drop}). 
Within this picture there is no counterpart of the 
rectangular-, trapezoidal-, and triangular-shaped barriers of the 
extended molecule model discussed in Sec.~\ref{sec-exp} enabling to intuitively 
define a transition voltage. However, 
it is still possible to find an analogy between the two models 
and to introduce a transition voltage $V_{t}^{p}$ for the point-like 
molecule as well. 

Since electron correlations are neglected in both models, one can cast the electric current 
expressed by means of the Landauer formula (see, e.~g.~\cite{HaugJauho}) in the following form 
%%%%%%%%%%%%%%%%%%%%%%%%%%%%%%%%%%%%%%%%%%%%%%%%%%%%%%%%%%%%%%%%%%%%%%%%%%%%%%%%%%%
\begin{eqnarray}
\displaystyle
I & = & I_S - I_D  , \nonumber \\
\label{eq-I}
I_S & = & 2\frac{e}{h}\int_{0}^{\infty} d\,\varepsilon T(\varepsilon, V) f(\varepsilon - \mu_S) , \\
I_D & = & 2\frac{e}{h}\int_{0}^{\infty} d\,\varepsilon T(\varepsilon, V) f(\varepsilon - \mu_D) ,
\nonumber
\end{eqnarray}
%%%%%%%%%%%%%%%%%%%%%%%%%%%%%%%%%%%%%%%%%%%%%%%%%%%%%%%%%%%%%%%%%%%%%%%%%%%%%%%%%%%
where $T$ and $f$ denote the transmission and the Fermi distribution function, 
respectively, and $h$ is Planck's constant. 
Above, the current $I$ has been split into the contribution $I_S$ of the electrons 
filling the levels in the source up to the energy $\mu_S=\varepsilon_F + eV/2$
that tunnel into the empty drain, and the contribution $I_D$ of the electrons
filling the levels in the drain up to the energy $\mu_D=\varepsilon_F - eV/2$
that tunnel into the empty source. 

At low source-drain 
voltages $V$ (the trapezoidal barrier of Fig.~\ref{fig:rect-trapez-triang}b), 
both contributions $I_S$ and $I_D$ are important, and the small difference 
between them gives rise to a small ohmic current $I \propto V$.
This mechanism is usually referred to as the direct tunneling 
\cite{Beebe:06,Choi:08,Choi:10}, but perhaps the term 
(nearly symmetric) bidirectional tunneling would be more suggestive.

If the source-drain voltage becomes sufficiently high, 
one arrives at a situation (the triangular barrier 
of Fig.~\ref{fig:rect-trapez-triang}c) where
the Fermi level of the drain lies below the bottom of the 
conduction band of the source.
Then, the electrons can only tunnel from the source to the drain but not in the 
reverse direction, since there are no final states available. This is the so-called
field emission mechanism (Fowler-Nordheim tunneling) 
\cite{Beebe:06,Choi:08,Chiu:08,Choi:10}, since it 
resembles the field emission of electrons from a metallic 
electrode. In fact, it is a unidirectional tunneling.

\begin{figure}[h]
% \centerline{\hspace*{-0ex}\includegraphics[width=0.3\textwidth,angle=0]{DirectTunneling.eps}}
% \centerline{\hspace*{-0ex}\includegraphics[width=0.3\textwidth,angle=0]{FieldEmission.eps}}
\centerline{\hspace*{-0ex}\includegraphics[width=0.8\textwidth,angle=0]{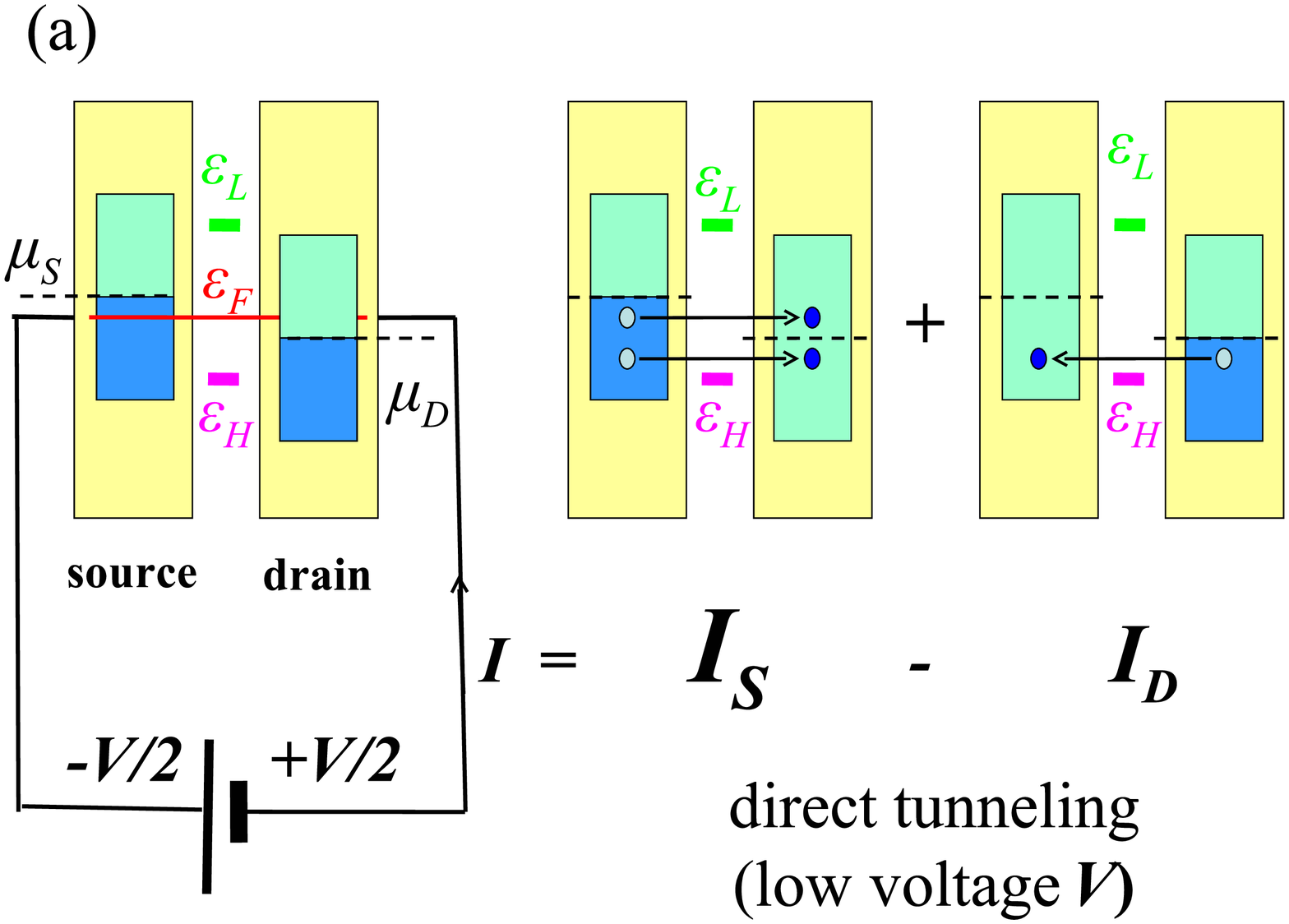}}
\centerline{\hspace*{-0ex}\includegraphics[width=0.8\textwidth,angle=0]{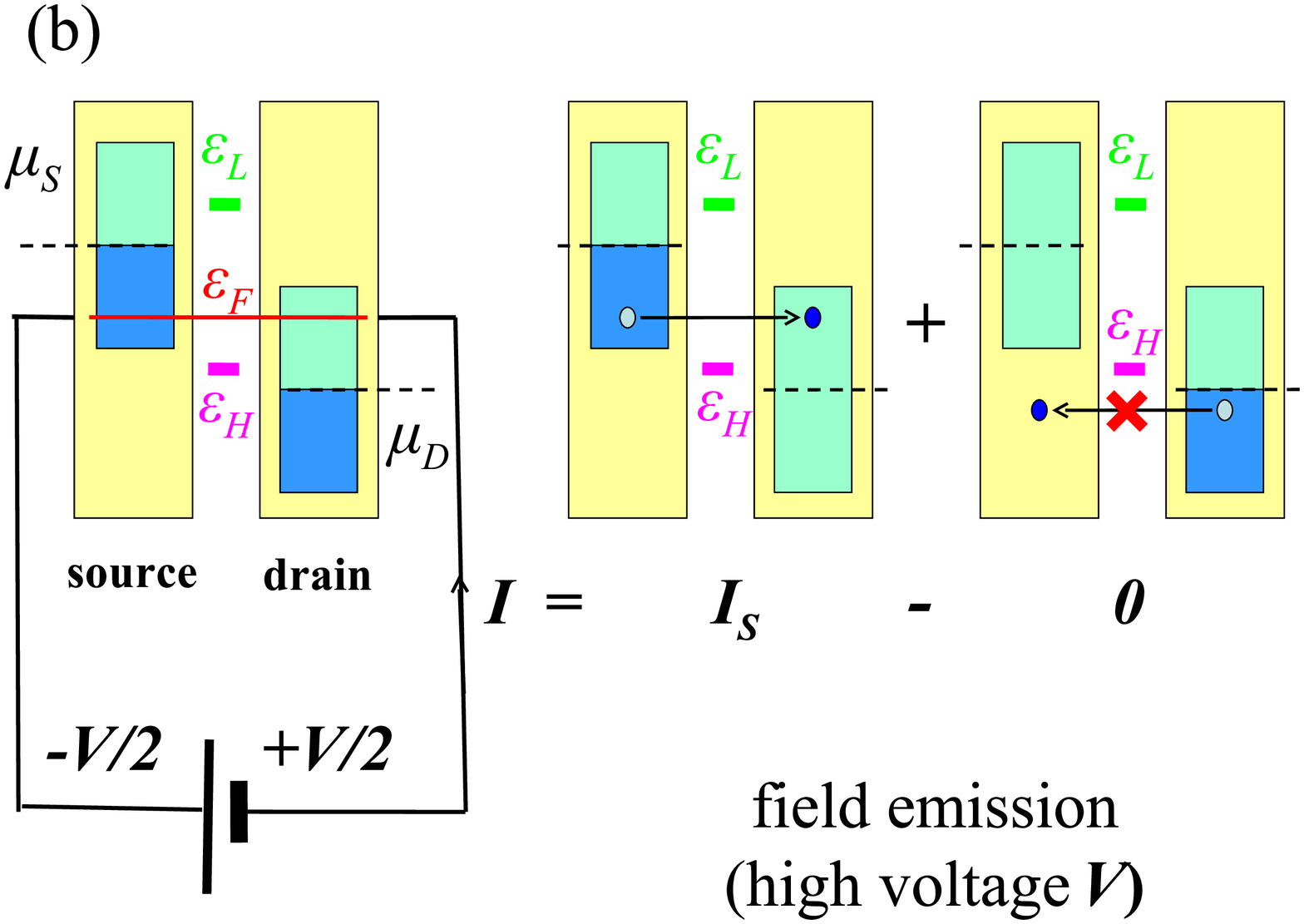}}
\caption{ The charge transfer mechanism 
is the bidirectional tunneling (direct tunneling) 
at lower source-drain voltages $V$ and unidirectional tunneling (field emission) at
higher voltages $V$.}
\label{fig:direct-tunn-field-emiss}
\end{figure}

While the idea of a voltage-driven change in the shape of the barrier 
(rectangular $\to$ trapezoidal $\to$ triangular) cannot be generalized 
from the homogeneously expanded molecule of Fig.~\ref{fig:rect-trapez-triang} 
to a point-like molecule, 
the notions of a tunneling that is symmetric or nearly bidirectional, 
highly asymmetric bidirectional or almost unidirectional, and 
completely unidirectional can be employed in both descriptions.
This is illustrated in Fig.~\ref{fig:direct-tunn-field-emiss}.  
Panels (a) and (b) of Fig.~\ref{fig:direct-tunn-field-emiss} 
shown for the case of a point-like molecule represent 
the counterpart of panels (b) and (c) of Fig.~\ref{fig:rect-trapez-triang} 
for an extended molecule. 
%%%%%%%%%%%%%%%%%%%%%%%%%%%%%%%%%%%%%%%%%%%%%%%%%%%%%%%%%%%%%%%%%%%%%%
%%%%%%%%%%%%%%%%%%%%%%%%%%%%%%%%%%%%%%%%%%%%%%%%%%%%%%%%%%%%%%%%%%%%%%

\begin{figure}[h]
% \centerline{\hspace*{-0ex}\includegraphics[width=0.3\textwidth,angle=0]{fig_FN_nrlm.eps}}
\centerline{\hspace*{-0ex}\includegraphics[width=0.8\textwidth,angle=0]{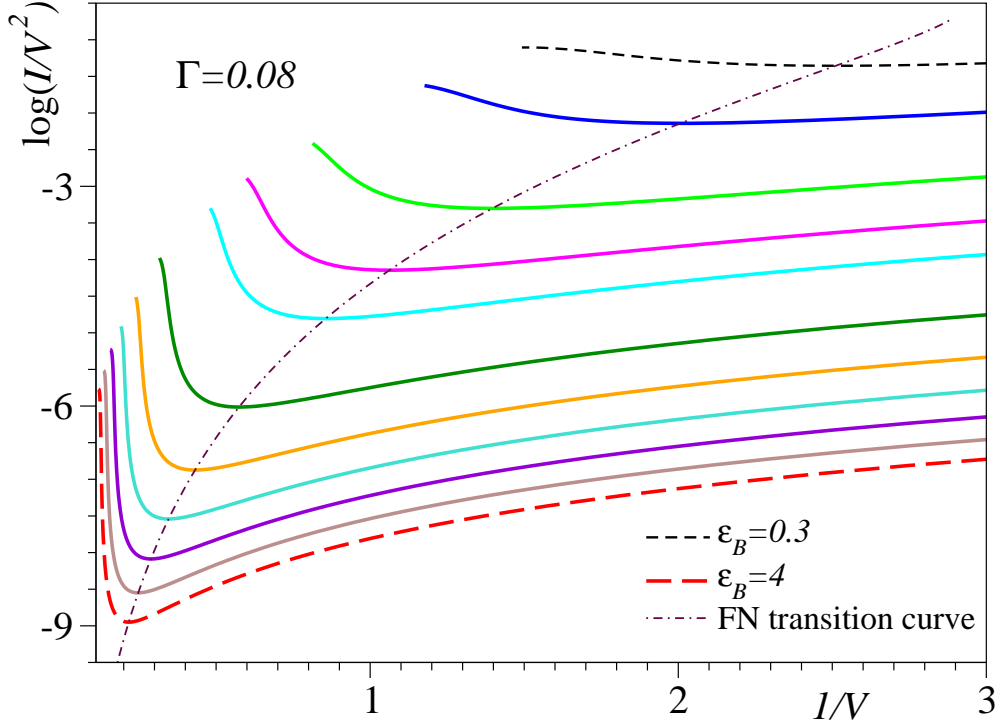}}
\caption{Fowler-Nordheim plots deduced from Eq.~(\ref{eq-wbl}) for the level energies 
$\varepsilon_{B}=0.3, 0.4, 0.6, 0.8, 1, 1.5, 2, 2.5, 3, 3.5, 4$ 
(values increase downwards). The dashed lines connecting the minima of the 
Fowler-Nordheim curves represents the Fowler-Nordheim transition line.}
\label{fig:FN-plots}
\end{figure}

The intuitive considerations presented above suggest that a transition between 
two different tunneling mechanisms can also be expected for a molecule
described within a point-like model. Within the common assumption of 
wide band electrodes, the current through a point-like level defined by the 
energy offset $\varepsilon_B$ can be expressed in a simple analytical form 
(see e.~g., \cite{HaugJauho,Baldea:2010e}) 
%%%%%%%%%%%%%%%%%%%%%%%%%%%%%%%%%%%%%%%%%%%%%%%%%%%%%%%%%%%%%%%%%%%%%%
\begin{equation}
\displaystyle
I = 2 \frac{e\Gamma}{h} % \frac{e}{h} \Gamma
\left(
\arctan\frac{e V -  2\varepsilon_B}{2\Gamma} + 
\arctan\frac{e V + 2\varepsilon_B}{2\Gamma}
\right) ,
\label{eq-wbl}
\end{equation}
%%%%%%%%%%%%%%%%%%%%%%%%%%%%%%%%%%%%%%%%%%%%%%%%%%%%%%%%%%%%%%%%%%%%%%
where $\Gamma$ represents the finite width of the molecular level 
due to the molecule-electrode coupling. In all numerical calculations, 
we set $e=h=1$ and the energy unit such that $\Gamma = 0.08$, but one should note
that by using the reduced variables 
$I/\Gamma$, $V/\Gamma$, and $\varepsilon_B/\Gamma$,
the $I$-$V$-characteristics expressed by Eq.~(\ref{eq-wbl})
are universal. 
By means of Eq.~(\ref{eq-wbl}),
one can straightforwardly obtain curves for 
$\log (I/V^2)$ plotted versus versus $1/V$ for various $\varepsilon_B$. 
These (Fowler-Nordheim) plots, which are shown in Fig.~\ref{fig:FN-plots},
clearly reveal a crossover between two regimes corresponding to low and high 
biases, which also exist for the point-like model, 
in agreement with the foregoing qualitative analysis. 
%%%%%%%%%%%%%%%%%%%%%%%%%%%%%%%%%%%%%%%%%%%%%%%%%%%%%%%%%%%%%%%%%%%%%%
%%%%%%%%%%%%%%%%%%%%%%%%%%%%%%%%%%%%%%%%%%%%%%%%%%%%%%%%%%%%%%%%%%%%%%

The Fowler-Nordheim transition line, which joins 
the minima \cite{inflexion} of the individual Fowler-Nordheim curves 
computed for various $\varepsilon_B$,
is presented in Fig.~\ref{fig:FN-plots}. The abscissae of this transition
line define the transition voltages $V_{t}^{p}$ between the two tunneling regimes.
The transition line of Fig.~\ref{fig:FN-plots} 
shows the parametrical dependence of $V_{t}^{p}$ on the energy offset 
$\varepsilon_B$. 
The curve expressing the explicit dependence $V_t^{p} = f(\varepsilon_B)$ is depicted in 
Fig.~\ref{fig:FN-critical-line}. By inspecting this curve, one can conclude that
this dependence is linear to a very good approximation
%%%%%%%%%%%%%%%%%%%%%%%%%%%%%%%%%%%%%%%%%%%%%%%%%%%%%%%%%%%%%%%%%%%%%%
\begin{equation}
\label{eq-Vt-p}
e V_{t}^{p} \simeq 1.15 \,\varepsilon_B 
\end{equation}
%%%%%%%%%%%%%%%%%%%%%%%%%%%%%%%%%%%%%%%%%%%%%%%%%%%%%%%%%%%%%%%%%%%%%%
provided that the energy offset $\varepsilon_B$ is sufficiently larger than the 
finite width $\Gamma$ caused by the molecule-electrode couplings.
%%%%%%%%%%%%%%%%%%%%%%%%%%%%%%%%%%%%%%%%%%%%%%%%%%%%%%%%%%%%%%%%%%%%%%%%%
\section{Discussion}
\label{sec-discussion}
Following the previous 
Refs.~\cite{Beebe:06,Choi:08,Choi:10,Reed:09}, I also employed 
above the notions of triangular barrier and field emission, but because they  
may be misleading, before discussing the aspects that are physically relevant, 
a few technical comments are in order.  

In the case of electrons emitted from a metal 
into the vacuum due to a strong electric field, 
the triangular barrier to be overcome and unidirectional 
tunneling \emph{concomitantly} exist. As discussed above,
both for a point-like molecule 
(cf.~Fig.~\ref{fig:direct-tunn-field-emiss}) and for an 
extended energy barrier \cite{Simmons:63}, the electron 
tunneling becomes unidirectional for rather high source-drain voltages, 
of the order of electrodes' bandwidth ($e V \sim 2\varepsilon_F$).
So, a transition from bidirectional tunneling to unidirectional tunneling 
occurs in both cases by sufficiently rising the voltage. The difference 
from the traditional vacuum electronics is that 
the ``triangular'' barrier regime encountered in molecular electronics 
does not \emph{necessarily} 
refer to a situation where the backward tunneling is impossible
but rather to situations where it is substantially inhibited.
As expressed by Eqs.~(\ref{eq-Vt}) and (\ref{eq-Vt-p}),
the system enters this regime at biases 
$e V \sim \varepsilon_B$, well in advance of reaching the ``field emission'' regime,
because normally $\varepsilon_B$ is substantially smaller than the electrodes' bandwidth.
For example, for the 
ODT-based junctions of Ref.~\cite{Reed:09}, $\varepsilon_B \simeq 1 - 2$\,eV,
and for the BDT-based junctions $\varepsilon_B \simeq 0.4 - 1.8$\,eV, while 
for the utilized gold electrodes $\varepsilon_F \simeq 5.6$\,eV.

For practical realizations in molecular electronics, it is important 
that the values of the energy offset $\varepsilon_B$
are sufficiently small, since too high values drastically suppress the current.
So, for such situations of practical interest, the transition voltage is not too high 
and, as witnessed by experiments \cite{Beebe:06,Choi:08,Chiu:08,Reed:09,Choi:10}, 
can be supported by already fabricated molecular junctions.
In particular, this justifies the use of the wide band approximation assumed above 
in Eq.~(\ref{eq-wbl}). Without this approximation 
the transmission function $T$ entering the 
Landauer formula, Eq.~(\ref{eq-I}), would exhibit a significant dependence on the applied 
bias $V$, and band edge effects could also become relevant even for a point-like molecule. 
Both these aspects have been discussed recently \cite{Baldea:2010e}.
In fact, as illustrated in Figs.~3a, 3b, and 4b of Ref.~\cite{Baldea:2010e}, 
the results obtained for a point-like molecule 
within and without the wide band approximation 
practically coincide up to voltages $e V \agt 2 \varepsilon_B$. 

As visible in the aforementioned figures of Ref.~\cite{Baldea:2010e}, 
the current rapidly increases at $e V \simeq e V_s \equiv 2 \varepsilon_B$. The physical reason
for this stepwise increase is that the frontier orbital becomes (nearly) resonant with the 
electrochemical potential of one electrode 
[$\mu_S(V_s) \equiv \varepsilon_F + eV_s/2 = \varepsilon_B + \varepsilon_F$ for the
n-type conduction supposed in the present figures]. Similar to $V_t$, 
this voltage $V_s$ is also straightforwardly related to the energy offset ($e V_s = 2 \varepsilon_B$);
it corresponds to an inflexion point of the $I$-$V$-characteristics \cite{I_S}.
This might suggest an alternative experimental procedure to determine 
$\varepsilon_B$ from $V_s$. While this method poses no problems for theoretical
calculations, it is unlikely that this method is experimentally viable: it is hardly conceivable 
that a molecular junction could support the high currents at voltages 
$V=V_s \simeq 2 V_t$ significantly higher than $V\simeq V_t$. 
By inspecting Fig.~1 and 2 of 
Ref.~\cite{Reed:09} as well as the present Fig.~\ref{fig:iv-nature}, 
one can remark that the high $V$-ranges sampled 
in the measurements comprise only small portions 
 ($V_s \gg V \agt V_t$) of the Fowler-Nordheim regime,
(presumably) just because the currents become too high.  

To conclude, what really happens at the transition voltage is that the frontier molecular 
orbital draws closer to the (electro)chemical potential of one 
electrode but still remains sufficiently distant from it. When the source-drain voltage
is raised, the point $V = V_t \approx \varepsilon_B$ is reached 
before reaching the point $V = V_s \approx 2 \varepsilon_B$.

Switching now to the physical aspects, I note the following.
The differences between the two models considered in 
Secs.~\ref{sec-exp} and \ref{sec-pcm} become notable 
only for nearly resonant 
frontier molecular orbitals $\varepsilon_B \sim \Gamma$. 
The fact that these differences are significant  
for $\varepsilon_B \sim \Gamma$ can easily be 
understood if one bears in mind that the semiclassical (WKB) 
calculations related to the model of an extended molecule 
do not account for finite width effects, which are important 
for the tunneling through a barrier of small height.

Definitely, the condition $\Gamma \ll \varepsilon_B$ 
holds in the experimental situation: by assuming a point-like 
molecule, from Fig.~S6 of the supplementary material 
of Ref.~\cite{Reed:09} one can estimate $\Gamma /\varepsilon_B = \sqrt{G/G_0} \sim 0.005$ 
for the ODT-based single molecule transistors, 
where $G$ is the zero-bias conductance and $G_0$ the conductance 
quantum.
So, the result of Eq.~(\ref{eq-Vt-p}), 
deduced for a point-like molecule, is not too much different from 
that obtained within the extended model of molecule, from the point where
the barrier shape changes from trapezoidal to triangular,
Eq.~(\ref{eq-Vt}). 
%%%%%%%%%%%%%%%%%%%%%%%%%%%%%%%%%%%%%%%%%%%%%%%%%%%%%%%%%%%%%%%%%%%%%%
\begin{figure}[h]
% \centerline{\hspace*{-0ex}\includegraphics[width=0.3\textwidth,angle=0]{fig_V_trans_nrlm.eps}}
\centerline{\hspace*{-0ex}\includegraphics[width=0.8\textwidth,angle=0]{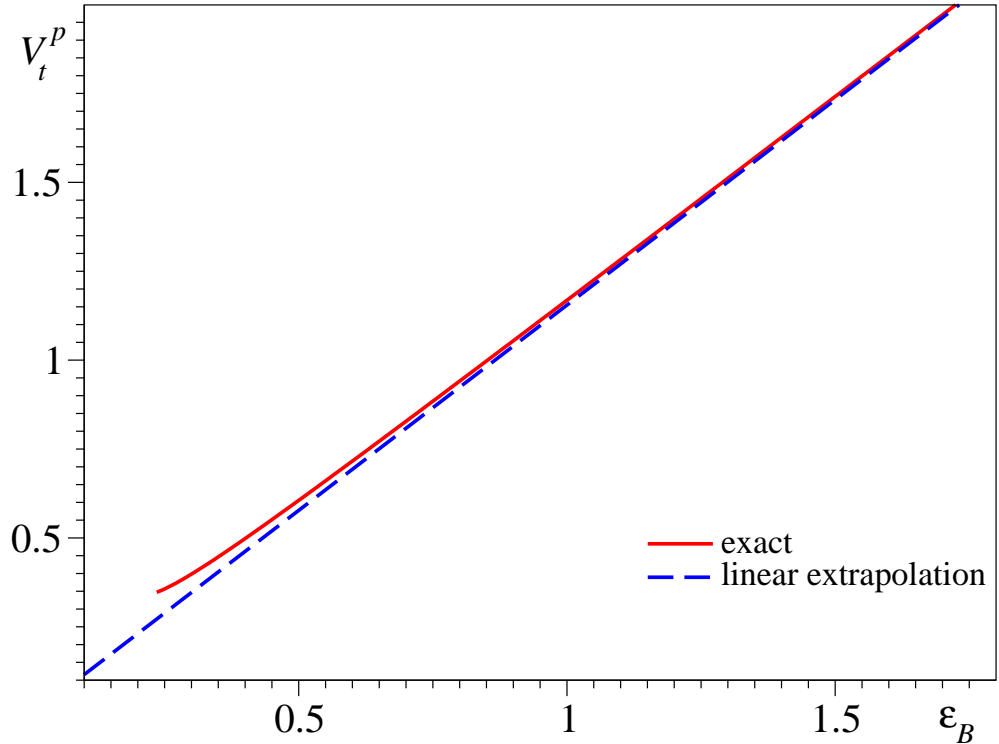}}
\caption{The Fowler-Nordheim critical line, which expresses the dependence of the
transition voltage on the energy offset $\varepsilon_B$ of the 
frontier molecular orbital, for a point-like molecule, deduced from
the minima of the Fowler-Nordheim plots of Fig.~\ref{fig:FN-plots}.}
\label{fig:FN-critical-line}
\end{figure}
%%%%%%%%%%%%%%%%%%%%%%%%%%%%%%%%%%%%%%%%%%%%%%%%%%%%%%%%%%%%%%%%%%%%%%

To summarize, for the realistic cases investigated in Ref.~\cite{Reed:09}
and elsewhere, which are  
characterized by frontier molecular orbitals that are 
sufficiently away from resonance ($\Gamma \ll \varepsilon_B$),
the dependence on the energy offset of the transition voltage 
of the point-like [$V_{t}^{p}$, Eq.~(\ref{eq-Vt-p})] 
and extended [$V_{t}^{e}$, Eq.~(\ref{eq-Vt})] models of molecule is linear. 
Therefore, the linear dependence of the transition voltage
(defined in conjunction with the Fowler-Nordheim plot) on the gate voltage
found in the experimental curves for current of Ref.~\cite{Reed:09} 
provides a clear evidence for the charge transfer through the 
gated HOMO whatever the model (expanded or point-like molecule) used to interpret 
the measurements. Employing the point-like model instead of the expanded 
one to interpret the experimental findings on the transition voltage of 
Ref.~\cite{Reed:09} only results in a slightly weaker coupling between 
the molecules and the gate electrode: instead of the values 
$\alpha = + 0.25$
and $\alpha = + 0.22$ estimated in Ref.~\cite{Reed:09} for the ODT 
and BDT  molecules, respectively, one obtains the slightly smaller 
conversion factors $\alpha = +0.22$ and $\alpha = + 0.19$, respectively.
The estimates of the two models agree within $\sim 13\%$, which is quite satisfactory 
if one bear in mind that the employed models, which are rather crude, 
do not account for many effects that could be significant, e.~g., 
possible potential drops at the contacts 
and charge image effects \cite{Simmons:63,Thygesen:09b}, as already noted 
\cite{Beebe:06}.

\begin{figure}[h]
% \centerline{\hspace*{-0ex}\includegraphics[width=0.3\textwidth,angle=0]{fig_log_i_log_v.eps}}
\centerline{\hspace*{-0ex}\includegraphics[width=0.8\textwidth,angle=0]{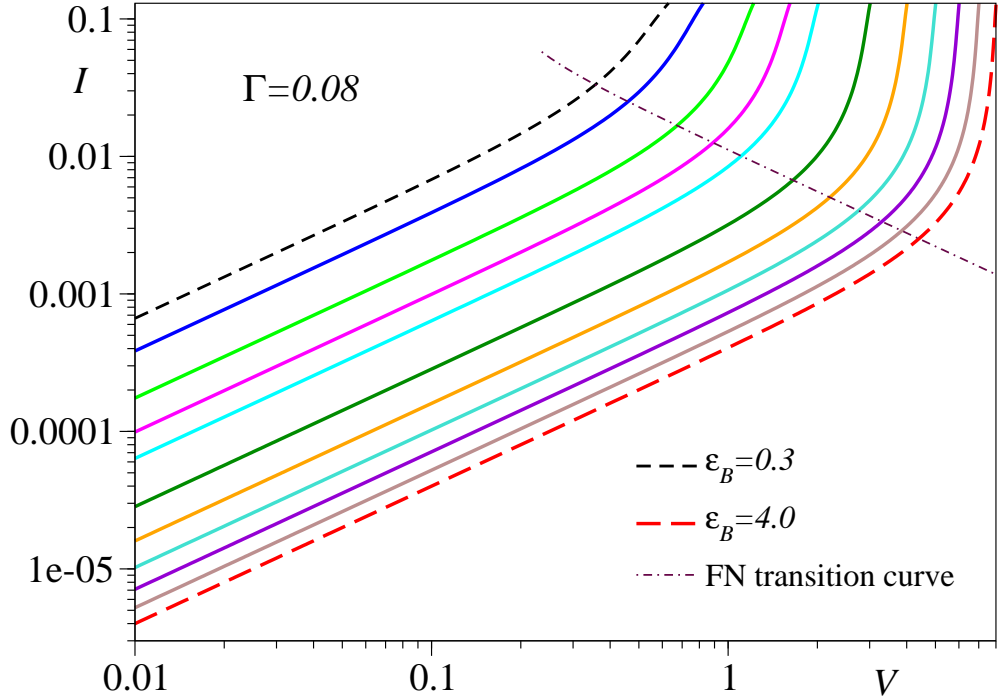}}
\caption{The $I-V$-characteristics of a point-like molecule 
[Eq.~(\ref{eq-wbl})]
plotted 
in logarithmic scale permit to evidence the crossover between 
two (low and high voltage) 
regimes and to relate it the Fowler-Nordheim transition line of 
Figs.~\ref{fig:FN-plots} and \ref{fig:FN-critical-line}.}
\label{fig:log-I-log-V}
\end{figure}

Interestingly, the dependence on $\varepsilon_B$ of the transition 
voltage deduced from the Fowler-Nordheim plot is linear, despite the fact that the 
underlying $I$-$V$ relationship of the two models 
[Eqs.~(\ref{eq-ohm}) and (\ref{eq-fn}) on one side, 
and Eq.~(\ref{eq-wbl}) on the other side] are quite different. 
This suggests that the examination of the Fowler-Nordheim plots 
could also be useful in other situations wherein the charge transfer 
mechanisms at lower and higher biases are different from 
those considered here. It is also interesting to mention 
that, similar to other cases \cite{Choi:08,Choi:10,frisbie}, the transition voltage
can also be visible in $\log I - \log V$ plots.
This becomes visible if one inspects Fig.~\ref{fig:log-I-log-V}, 
where such plots are presented along 
with the Fowler-Nordheim transition line depicted in 
Figs.~\ref{fig:FN-plots} and \ref{fig:FN-critical-line}.
Noteworthy, unlike the plots of Figs.~\ref{fig:FN-plots} and \ref{fig:log-I-log-V}, 
the $I-V$-characteristics represented in the usual 
linear scale do not exhibit any special feature at the 
transition voltage $V=V_t$, and therefore are not shown here.
%%%%%%%%%%%%%%%%%%%%%%%%%%%%%%%%%%%%%%%%%%%%%%%%%%%%%%%%%%%%%%%%%%%%%%
%%%%%%%%%%%%%%%%%%%%%%%%%%%%%%%%%%%%%%%%%%%%%%%%%%%%%%%%%%%%%%%%%%%%%%
\section{Conclusion}
\label{sec-conclusion}
Remarkably, whether the molecule is modeled by a energy barrier 
continuously expanded  
in the space between the source and the drain, or by a point-like energy 
level, the transition voltage $V_t$ extracted as described above
exhibits a \emph{linear} dependence 
on the energy offset between the frontier molecular orbital and the 
electrodes' Fermi energy. 
The difference between these two extreme descriptions, which  
only amounts $\sim 13$\% is quite reasonable in view of the fact that 
both descriptions represent rather crude approximations of a realistic 
molecular junction. 
Because it is plausible to assume that the potential profile in a 
real molecule corresponds to 
a situation, which lies between the situations described by these 
two extreme physical models, one can conclude that the evidence on molecular orbital 
gating reported in Ref.~\cite{Reed:09} is not restricted to the assumption 
made of the change in the barrier shape from trapezoidal to triangular and 
also holds in a model that is completely different. 

Of course, the present work 
only represents a preliminary investigation of the molecular orbital gating, 
as it is based on a model, which is as oversimplified as the barrier model employed in 
Ref.~\cite{Reed:09}. To 
quantitatively reproduce the $I$-$V$-characteristics 
measured in the single-molecule transistors of Ref.~\cite{Reed:09}
subsequent investigations should reliably account for effects 
that have to be considered in real systems. 
Open challenge for future investigations are,
e.~g., the inclusion of the chemical information about the molecule, electron correlation 
effects, and selfconsistent determination of the potential distribution,
as well as the numerous affinity and ionization levels of a real molecule, 
which should be used instead of the LUMO and the HOMO, respectively.
The latter have been employed in the present paper to keep the discussion as simple as 
that of Ref.~\cite{Reed:09}, but this should not create the wrong impression that a real experiment
can actually be interpreted in terms of a single molecular orbital.
%%%%%%%%%%%%%%%%%%%%%%%%%%%%%%%%%%%%%%%%%%%%%%%%%%%%%%%%%%%%%%%%%%%%%%
\section*{Acknowledgments}
%%%%%%%%%%%%%%%%%%%%%%%%%%%%%%%%%%%%%%%%%%%%%%%%%%%%%%%%%%%%%%%%%%%%%%
The author is very pleased to thank Horst K\"oppel for numerous discussions 
and a fruitful collaboration over many years.
He also thanks Dan Frisbie for useful clarifications on the 
experimental determination of the transition voltage. 
Financial support for this work 
provided by the Deu\-tsche For\-schungs\-ge\-mein\-schaft (DFG) is gratefully 
acknowledged.
%%%%%%%%%%%%%%%%%%%%%%%%%%%%%%%%%%%%%%%%%%%%%%%%%%%%%%%%%%%%%%%%%%%%%%%%%%%%%%%%%%%%%%%%%%%%
%%%\bibliographystyle{model1a-num-names} % USED FOR CHEM. PHYS.
% \bibliographystyle{model1-num-names} % pp. title
% \bibliographystyle{model1b-num-names}  % pp. title
% \bibliographystyle{model1c-num-names} % pp.
% \bibliographystyle{model2-names} % PRB + url + doi
% \bibliographystyle{model3a-num-names} % pp. title url doi
% \bibliographystyle{model3-num-names} % PRB + url + doi
% \bibliographystyle{model4-names} % PRB + url + doi
% \bibliographystyle{model5-names} % PRB + url + doi
% \bibliographystyle{model6-num-names} % chemical style year:vol:p
% \bibliographystyle{elsarticle-harv} % pp. title
% \bibliography{/home/ioan/QDs/LinearResponse/paper/bibl} 

\begin{thebibliography}{24}
\bibitem[{sym(????)}]{symmetrical-location}
\bibinfo{note}{From the fact that the $I$-$V$-characteristics of Figs.~1a and 2a 
of Ref.~\cite{Reed:09} are invariant with respect to the transformation $(I\to -I, V\to -V)$
one can indeed infer that the molecule is located symmetrically between the source and the drain.}
\end{thebibliography}
% \end{document}
%%%%%%%%%%%%%%%%%%%%%%%%%%%%%%%%%%%%%%%%%%%%%%%%%%%%%%%%%%%%%%%%%%%%%%%%%%%%%%%%%%%%%%%%%%%%

%%%%%%%%%%%%%%%%%%%%%%%%%%%%%%%%%%%%%%%%%%%%%%%%%%%%%%%%%%%%%%%%%%%%%%%%%%%%%%%%%%%%%%%%%%%%
\end{document}